\documentclass[submission,copyright,creativecommons]{eptcs}
\providecommand{\event}{GandALF 2012} % Name of the event you are submitting to
\usepackage{breakurl}             % Not needed if you use pdflatex only.

\usepackage{amsmath}
\usepackage{amssymb}
\usepackage{graphicx}

\newcommand{\tabularblock}[3]{\begin{tabular}{l}#1\\#2 \vspace{#3}\end{tabular}}
\newcommand{\tabularblocksingleton}[1]{\begin{tabular}{l}#1\end{tabular}}

\newcommand{\arrowhead}{\!\!\!\!\!
\scalebox{0.5}{\begin{tabular}{l}\vspace{0.9mm}\textgreater\vspace{0.9mm}\end{tabular}}
\,
}

\newcommand{\transition}[3]{#1\,\,\frac{\enspace #3 \enspace}{\text{\enspace}}
\arrowhead
#2}
\newcommand{\transitionf}[3]{#1\,\,\frac{\enspace #3 \enspace}{\enspace \text{F} \enspace}\arrowhead #2}
\newcommand{\longtransition}[5]
{
#1\,\,\frac{\enspace #4 \enspace}{\text{\enspace}}\arrowhead #2\,\,\frac{\enspace #5 \enspace}{\text{\enspace}}\arrowhead #3
}

\newcommand{\longtransitionf}[5]
{
#1\,\,\frac{\enspace #4 \enspace}{\bullet}\arrowhead #2\,\,\frac{\enspace #5 \enspace}{\bullet}\arrowhead #3
}

\newcommand{\tmin}{\text{min}}

\newcommand{\emptyword}{\varepsilon}

\title{Learn with SAT to Minimize B\"uchi Automata}

\author{
Stephan Barth
\institute{Ludwig-Maximilians-Universit\"at M\"unchen, Germany}
\and Martin Hofmann
\institute{Ludwig-Maximilians-Universit\"at M\"unchen, Germany}
}

\def\titlerunning{Minimize B\"uchi Automata}
\def\authorrunning{S. Barth \& M. Hofmann}
\begin{document}
\maketitle

\begin{abstract}
We describe a minimization procedure for nondeterministic B\"uchi
automata (NBA).
For an automaton $A$ another automaton $A_{\min}$ with the minimal
number of states is learned with the help of a SAT-solver.

This is done by successively computing automata $A'$ that
approximate $A$ in the sense that they accept a given
finite set of positive examples and reject a given
finite set of negative examples. In the course of the procedure these
example sets are successively increased. Thus, our method can be seen
as an instance of a generic learning algorithm based on a ``minimally adequate teacher'' in the sense of Angluin.

We use a SAT solver to find an NBA for given sets of positive and
negative examples. We use complementation via construction of
deterministic parity automata to check
candidates computed in this manner for equivalence with $A$.  Failure
of equivalence yields new positive or negative examples. Our method
proved successful on complete samplings of small automata and
of quite some examples of bigger automata.

We successfully ran the minimization on over ten thousand automata
with mostly up to ten states, including the complements of all possible
automata with two states and alphabet size three and discuss
results and runtimes; single examples had over 100 states.

%In particular, we succeed in minimizing a big percentage of NBA with
%10 states as well as all NBA and their complements with two states
%and two- and three-letter alphabet as well as some instances of
%Michel's NBA which were introduced to establish an $n!$
%lower bound on complementation of NBA.

%\keywords{B\"uchi Automata, SAT solver, Minimization, Automata learning}
\end{abstract}

\section{Introduction}
Minimization is a well-studied and widely used principle in
many areas. In the theory of automata the  best known
example is the minimization of deterministic finite automata (DFA).
It has the interesting property that by using only local optimizations
one will always reach the same global minimum.
This property is not valid anymore for some other automata models,
nevertheless local optimization can still achieve a considerable reduction
in size.

Because of that and its applications in automatic verification and
other fields some incomplete minimization algorithms of
nondeterministic B\"uchi automata (NBA) have been studied.  They
include local (\cite{optimizing} p. 6--11) minimizations, and other
minimizations that do not guarantee to find a smallest automaton but
only reduce the size \cite{spin2010}.  Other studied minimization
algorithms only work on some kind of B\"uchi automata (deterministic
B\"uchi automata \cite{sat2010} or deterministic weak B\"uchi automata
\cite{loeding}).

These algorithm try to balance computational efficiency
with low size of the resulting NBA or with generality.
%While all these limitations reduce runtime significantly, they are
%still limitations.
After application of these algorithms it is not
guaranteed that a found automaton is minimal nor can minimality of a
given automaton be proven, or they are not applicable to all automata.

While this status is sufficient for many applications it is unsatisfactory
not to have any algorithms for global minimization of NBA; on the
theoretical side it is a gap, on the practical side it means that one never
knows whether a given automaton might admit further reduction in size;
especially when representing a policy the used automata are often very
small and every additional state increases the resource consumption
noticeable.

We present here the first procedure that computes for a given B\"uchi
automaton an equivalent one of minimal size among all B\"uchi automata
equivalent to the given one. We call this ``global minimization for
B\"uchi automata''.
 
Unlike in the case of deterministic finite automata such minimal
automata are, however, not unique up to isomorphism.

Our approach can be seen as an instance of Angluin's learning
framework and indeed would be able to construct a minimal NBA for an
arbitrary $\omega$-regular language presented by a minimally adequate
teacher in the sense of \cite{angluin87}.

\subsection{B\"uchi Automata}
A nondeterministic B\"uchi automaton (NBA) describes a language
of infinite words. It is given by a tuple $(Q, \Sigma, q_0, F, \delta)$
where $Q$ is a finite set of states, $\Sigma$ a finite alphabet, $q_0 \in Q$ the
starting state, $F \subset Q$ the set of final states, and
$\delta: Q\times \Sigma \rightarrow 2^Q$ the transition function. A word
$a_0a_1a_2\dots \in \Sigma^\omega$ is said to be accepted if and only if
$\exists q_1q_2q_3\dots$ such that $\forall i\in {\mathbb N}_0. q_{i+1}\in
\delta(q_i, a_i)$ and $\forall i\in {\mathbb N}\, \exists j>i. q_j \in F$.

\begin{figure}[h]
\includegraphics[viewport=-290 30 -270 40,height=1.5mm]{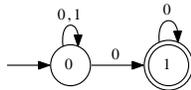}
\caption{Example NBA, accepting the language $(0|1)^*0^\omega$
\label{nbaexample}
}
\end{figure}

%\begin{tabular}{p{7.8cm}l}
For example, a B\"uchi automaton for the language
$L=(0|1)^*0^\omega$
(``finitely many 1s'') is shown in Figure~\ref{nbaexample}
% &

%\end{tabular}

A run $q_0q_1q_2\dots$ on this automaton for a word $w\in L$
is obtained\\
by choosing
$k \in {\mathbb N}.\forall l > k.w_l=0$ and setting $0=q_0=\dots=q_k$
and $1=q_{k+1}=\dots$

%One defines deterministic B\"uchi automata (DBA) by requiring that the
%transition function always yields a singleton set. It is easy to see
%that the previous language cannot be recognized by a DBA.

One defines the $\omega$-regular languages as those recognized by
NBA.
%It is a standard result that every $\omega$-regular language can
%be written in the form $\bigcup_{i=1}^{n}U_iV_i^{\omega}$ for regular
%(in the standard, finite-word sense) languages $U_i, V_i$. Herein,
%$UV^\omega=\{uv_1v_2v_3\dots\mid u\in U, v_i\in V\}$.

\subsection{Problem complexity}
DFA can be minimized in
polynomial time \cite{Hopcroft_1971} whereas minimization of
deterministic B\"uchi automata is
%already
NP-complete \cite{CoRR10,sat2010}.

In case of NBA, the minimization problem is PSPACE-complete as it is
already PSPACE-complete for nondeterministic finite automata
%(even
%calculation of the minimal size is PSPACE-complete
(\cite{Gramdis} page
27, theorem 3) and it is easy to see that minimization of B\"uchi
automata is in PSPACE given the well-known fact that equivalence of
NBA is in PSPACE.

This in itself is not necessarily a problem because results of
absolute minimization are nontrivial and of interest even for small
problem instances. One may also remark that there exist practically
and even industrially successful implementations of PSPACE hard
problems, consider e.g.\ LTL model checking as implemented in the SPIN
tool \cite{Holzmann:2003:SMC:1405716} or even the WMSO implementation
MONA \cite{monamanual2001}.

The minimization procedure presented still leaves scope for further
optimization, yet it is able to produce nontrivial and hitherto unknown
results. For example, we were able to ascertain that in case of
a two letter alphabet the complements of
B\"uchi automata with two states require at most five states; we were
also able to assert the minimality of the first instances of
Michel's family of NBA \cite{Michel88}, the first member of this family has two letters
and two states and needs five states for its complement thus matching the
here found limit for complement size for this automata size.

\subsection{SAT solver}

The abovementioned minimization algorithm of DBA \cite{sat2010} uses a
SAT-solver to search for a DBA equivalent to a given one with a
smaller number of states. To this end, equivalence of automata is
encoded directly as a SAT formula which is possible since equivalence
of DBA is in P. Since equivalence of NBA is PSPACE complete, this approach
does not extend to NBA directly; nevertheless a SAT solver is a useful
tool in our approach.

A SAT solver is a software that takes a boolean formula in conjunctive
normal form (CNF) presented as a list of clauses in some machine
readable format and returns a satisfying assignment if the formula is
satisfiable and answers ``unsatisfiable'' otherwise.

Although satisfiability of CNF is NP-complete, modern SAT
solvers can be applied to practically relevant and appreciably large
instances. On modern computers, instances with 1000 variables
and 10000 clauses are
solvable in reasonable time. In specific cases even larger instances
are solvable. This has earned SAT solvers a tremendous and still
increasing popularity in recent years.

%While the standard construction of CNF result in exponential bigger
%formulas introduction of fresh variables can limitate this blowup to
%polynomial size.
While the standard construction of CNF results in exponentially bigger
formulas, introduction of fresh variables can limit this blowup to
polynomial size.

\section{Overview over the algorithm}

The original automaton is transformed into a teacher for NBA
in sense of Angluin \cite{angluin87} by performing equivalence
tests for constructing counterexamples or returning true.

The core is a candidate finder that creates B\"uchi automata
out of positive (called good words) and negative (called bad words) word
examples and additionally ensures minimal size for automata classifying
these examples.

This is used to find candidates for the minimal
automaton.
%Checking these candidates results in a positive
%result and thus a minimal automaton or a negative result and generates
%new good or bad words.
A candidate is checked against the original automaton.
In case of equivalence the candidate is a minimum
automaton, whereas inequivality results in new good or bad words.

The candidate finder is presented in section \ref{candidatefinder}, the
algorithm using the learner as black box in
section \ref{algorithm}. Pseudo code presenting both at once is given
in Figure~\ref{pseudocode}.

\subsection{Notation}

The following notations are used in this paper:
\begin{itemize}
\item $w_i$ denotes the $i$-th letter of the word $w$.
\item
$[a]$ denotes the one-letter word consisting of $a\in\Sigma$;
\item
$\transition{i}{j}{w}$ denotes a transition with the word $w$ from
state $i$ to state $j$;
\item
$\transitionf{i}{j}{w}$ denotes a transition with the word $w$ from
state $i$ to state $j$ with a visit of a final state anywhere on this path
(including $i$ and $j$);
\item
$\longtransition{i}{j}{k}{v}{w}$ is short for
$\transition{i}{j}{v} \land \transition{j}{k}{w}$;
\item
$\longtransitionf{i}{j}{k}{v}{w}$ is short for
$(\transition{i}{j}{v} \land \transitionf{j}{k}{w}) \lor (\transitionf{i}{j}{v} \land \transition{j}{k}{w})$.
\item For an automaton $A$ we denote the language of the automaton by $L(A)$.
\end{itemize}

\subsection{Candidate finder for B\"uchi automata}
\label{candidatefinder}
The candidate finder generates from given finite sets $G$
and $B$ of ultimately periodic words and an integer
value $n$ a SAT formula whose satisfying assignments
precisely correspond to automata $A'$ with $n$ states
such that $G \subseteq L(A') \subseteq \overline{B}$.

%The candidate finder generates a NBA out of sets of positive and negative
%examples; these sets are called good and bad words. The good words $G$
%have the property that they should be in the language of the automaton
%searched for, the bad words $B$ should be not in the language of the
%automaton searched for. Any candidate $A'$ such
%that $G \subset L(A') \subset \overline{B}$ is herby a solution.

%For achieving this the words are encoded in a SAT formula.
The SAT formula represents an unknown automaton $X$ with $n$ states
using variables $t_{i,j,a}$ (an $a$-labelled edge from $i$ to $j$) and
$f_i$ (finality of state $i$).
%. Furthermore, we use
Further variables are defined, including $z_{u,v}$ ($uv^\omega$ is accepted by $X$).
The formula itself
then has to ensure these intended meanings and additionally
comprises the conjunction of $z_{u,v}$ for $uv^\omega\in G$
and $\neg z_{u,v}$ for $uv^\omega\in B$.

\begin{table}[h]
\centering
\begin{tabular}{l|p{32em}}
variable & meaning\\% & count\\
\hline
\hline
$f_{i}$ & State $i$ is final state\\% & $\Theta(n)$\\
\hline
$t_{i,j,a}$ & $\transition{i}{j}{[a]}$\\% & $\Theta(n^2 \cdot \sigma)$\\
\hline
$d_{i,j,w}$ & $\transition{i}{j}{w}$\\% & $\Theta(n^2 \cdot \rho)$\\
\hline
$o_{i,j,k,a,w}$ & $\longtransition{i}{j}{k}{[a]}{w}$\\% & $\Theta(n^3 \cdot \rho)$\\
\hline
$x_{w,i,j,m}$ & There is a $k\in \{1,\dots, 2^m\}$, such that
$\transition{i}{j}{w^k}$
\\%& $\Theta(\kappa \cdot n^2 \cdot \log n)$\\
\hline
$h_{w,i,j,k,m}$ & There are $l_1,l_2\in \{1,\dots,2^m\}$, such that
$\longtransition{i}{j}{k}{w^{l_1}}{w^{l_2}}$
\\%& $\Theta(\kappa \cdot n^3 \cdot \log n)$\\
\hline
$D_{i,j,w}$ &
$\transitionf{i}{j}{w}$
\\%& $\Theta(n^2 \cdot \rho)$\\
\hline
$O_{i,j,k,a,w}$ &
$\longtransitionf{i}{j}{k}{[a]}{w}$
\\%& $\Theta(n^3 \cdot \rho)$\\
\hline
$s_{u,v,i,m}$ &
There is a $k \in \{1,\dots,2^m\}$ that
$\transition{q}{i}{uv^k}$
\\%& $\Theta(\kappa \cdot n \cdot \log n)$\\
\hline
$u_{u,v,i,j,m}$ &
There is a $k \in \{1,\dots,{2^m}\}$ that
$\longtransition{q}{i}{j}{u}{v^k}$
\\%& $\Theta(\kappa \cdot n^2 \cdot \log n)$\\
\hline
$B_{i,j,w,m}$ & There is a number $k \in \{1,\dots,2^m\}$ such that
$\transitionf{i}{j}{w} \land \transition{j}{i}{w^k}$
\\%& $\Theta(\kappa \cdot n^2 \cdot \log n)$\\
\hline
$L_{i,w,m}$ & There is a number $k \in \{1,\dots,2^m\}$ and a state $j$
such that
$\transitionf{i}{j}{w} \land \transition{j}{i}{w^k}$
\\%& $\Theta(\kappa \cdot n \cdot \log n)$\\
\hline
$y_{u,v,i}$ &
There are $k_1,k_2 \in \{1,\dots, 2^{[\log_2(n)]+1}\}$ such that
$\transition{q}{i}{uv^{k_1}} \land \transitionf{i}{i}{v^{k_2+1}}$ (is $uv^\omega$ accepted via the state $i$ as loop knot).
\\%& $\Theta(\kappa \cdot n)$\\
\hline
$z_{u,v}$ & The word $uv^\omega$ is accepted.\\% & $\Theta(\kappa)$\\
\end{tabular}
\caption{Variables used in the SAT encoding
\label{variablesinsatenc}
}
\end{table}

\begin{table}[h]
\centering
\begin{tabular}{l|l}
variable & deduction\\
\hline
$d_{i,j,\emptyword}$ & $i=j$\\
%\hline
$d_{i,j,[a]}$ & $t_{i,j,a}$\\
%\hline
$d_{i,j,a.w}$ & $\bigvee_{k=0\dots n-1} o_{i,j,k,a,w}$\\
\hline
$o_{i,j,k,a,w}$ & $d_{i,k,[a]}\land d_{k,j,w}$\\
\hline
$x_{\emptyword,i,j,m}$ & $i=j$\\
%\hline
$x_{w,i,j,0}$ & $d_{i,j,w}$\\
%\hline
$x_{w,i,j,m}$ & $x_{w,i,j,m-1} \lor \bigvee_{k=0\dots n-1}h_{w,i,k,j,m-1}$\\
\hline
%$h_{w,i,j,k,m}$ & $x_{w,i,j,m} \land x_{w,j,k,m}$\\
%\hline
$z_{u,v}$ & $\bigvee_{k=0\dots n-1} y_{u,v,k}$
\end{tabular}
\caption{Definition of variables (selection)
\label{variabledeductions}
}
\end{table}

Table~\ref{variablesinsatenc} summarises the variables used in the expression.
The variables are chosen in a way that every variable can be deduced
by a small (constant size or linear in count of states) SAT formula
from other variables; this limits the blowup for generating a CNF to
polynomial instead of exponential size. These deductions follow in a
simple way from their meaning; for some
variables these deductions are shown in Table~\ref{variabledeductions}.

All in all the SAT expression consists of linear many variables as function of
the alphabet size, the number of good and bad words and the length
of the good and bad words. There are cubic many as function of
the size of the automaton searched for.

\begin{figure}[h]
%The external visibly variables of this expression are:
The external used variables of this expression are:
\begin{itemize}
\item $z_{"","0"}$ (acceptance of the word $0^\omega$),
\item $z_{"","1"}$ (acceptance of the word $1^\omega$),
\item $t_{0,0,'0'}$ (existence of transition with letter $0$ from state $0$ to state $0$),
\item $t_{0,0,'1'}$ (existence of transition with letter $1$ from state $0$ to state $0$) and
\item $f_{0}$ (finality of state $0$).
\end{itemize}
{\small
$(u_{"","1",0,0,1} \iff (d_{0,0,""} \land x_{"1",0,0,1})) \land
(s_{"","1",0,1} \iff u_{"","1",0,0,1}) \land
(D_{0,0,"1"} \iff ({\bf t_{0,0,'1'}} \land {\bf f_{0}})) \land
(d_{0,0,"1"} \iff {\bf t_{0,0,'1'}}) \land
(x_{"1",0,0,0} \iff d_{0,0,"1"}) \land
(h_{"1",0,0,0,0} \iff x_{"1",0,0,0}) \land
(x_{"1",0,0,1} \iff (x_{"1",0,0,0} \lor h_{"1",0,0,0,0})) \land
(B_{0,0,"1",1} \iff (D_{0,0,"1"} \land x_{"1",0,0,1})) \land
(L_{0,"1",1} \iff B_{0,0,"1",1}) \land
(y_{"","1",0} \iff (s_{"","1",0,1} \land L_{0,"1",1})) \land
({\bf z_{"","1"}} \iff y_{"","1",0}) \land
(d_{0,0,""}) \land
(u_{"","0",0,0,1} \iff (d_{0,0,""} \land x_{"0",0,0,1})) \land
(s_{"","0",0,1} \iff u_{"","0",0,0,1}) \land
(D_{0,0,"0"} \iff ({\bf t_{0,0,'0'}} \land {\bf f_{0}})) \land
(d_{0,0,"0"} \iff {\bf t_{0,0,'0'}}) \land
(x_{"0",0,0,0} \iff d_{0,0,"0"}) \land
(h_{"0",0,0,0,0} \iff x_{"0",0,0,0}) \land
(x_{"0",0,0,1} \iff (x_{"0",0,0,0} \lor h_{"0",0,0,0,0})) \land
(B_{0,0,"0",1} \iff (D_{0,0,"0"} \land x_{"0",0,0,1})) \land
(L_{0,"0",1} \iff B_{0,0,"0",1}) \land
(y_{"","0",0} \iff (s_{"","0",0,1} \land L_{0,"0",1})) \land
({\bf z_{"","0"}} \iff y_{"","0",0}) \land
({\bf z_{"","1"}}) \land
(\neg {\bf z_{"","0"}})
$
}

Solution computed by Minisat:

$u_{"","1",0,0,1}$, $d_{0,0,""}$, $x_{"1",0,0,1}$, $s_{"","1",0,1}$, $D_{0,0,"1"}$, ${\bf t_{0,0,'1'}}$, ${\bf f_{0}}$, $d_{0,0,"1"}$, $x_{"1",0,0,0}$, $h_{"1",0,0,0,0}$, $B_{0,0,"1",1}$, $L_{0,"1",1}$, $y_{"","1",0}$, ${\bf z_{"","1"}}$, $\neg u_{"","0",0,0,1}$, $\neg x_{"0",0,0,1}$, $\neg s_{"","0",0,1}$, $\neg D_{0,0,"0"}$, $\neg {\bf t_{0,0,'0'}}$, $\neg d_{0,0,"0"}$, $\neg x_{"0",0,0,0}$, $\neg h_{"0",0,0,0,0}$, $\neg B_{0,0,"0",1}$, $\neg L_{0,"0",1}$, $\neg y_{"","0",0}$, $\neg {\bf z_{"","0"}}
$

\caption{SAT expression for $G=\{1^\omega\}$, $B=\{0^\omega\}$, $n=1$ and its solution
\label{exampleexpression}
}
\end{figure}

For further illustration, we present in Figure~\ref{exampleexpression}
the entire expression corresponding to $G=\{1^\omega\}$ and
$B=\{0^\omega\}$ and $n=1$ as well as its satisfying assignment
computed by Minisat. For better readability the formula is presented
not in CNF while it is in CNF in the implementation. The only needed
transformation for creating CNF is resolving the equivalences
(denoted by ``$\iff$'') thus roughly doubling the size of the expression.

\begin{table}[h]
\centering
\begin{tabular}{l|l||l|l}%|p{22mm}}
example words & resulting automaton &
example words & resulting automaton\\% & Anmerkung\\
\hline
\hline
\tabularblock{$G=\{1^\omega\}$}{$B=\{0^\omega\}$}{2mm} &
\tabularblocksingleton{\includegraphics[viewport=50 0 150 90,height=10.8mm]{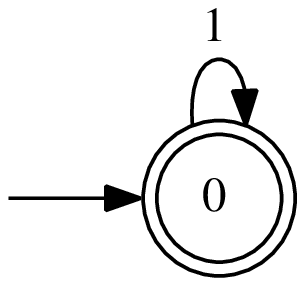}} &
\tabularblock{$G=\{0^\omega, 1^\omega\}$}{$B=\{(01)^\omega\}$}{2mm} &
\tabularblocksingleton{\includegraphics[viewport=50 0 240 90,height=10.8mm]{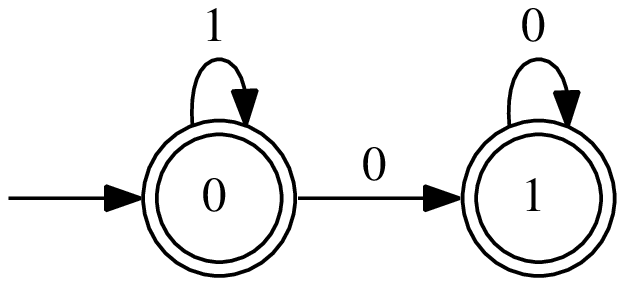}} \\
\hline
\tabularblock{$G=\{01^\omega, 10^\omega\}$}{$B=\{0^\omega, 1^\omega\}$}{6mm} &
\tabularblocksingleton{\includegraphics[viewport=50 0 235 190,height=22.8mm]{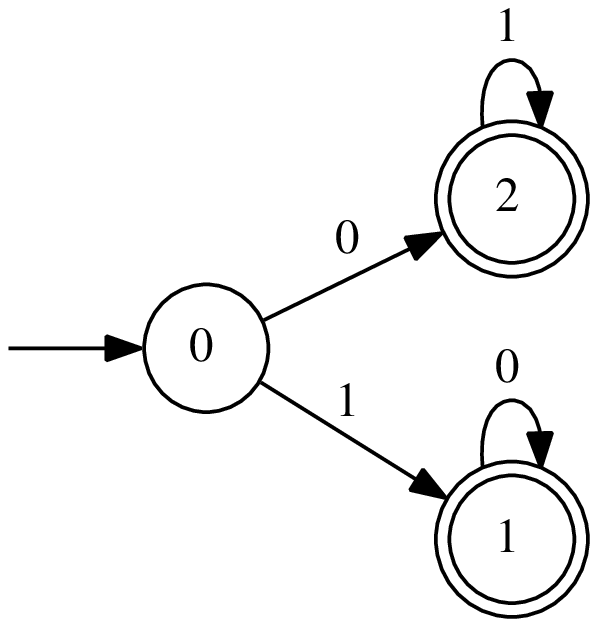}} &
\tabularblock{$G=\{(01)^\omega\}$}{$B=\{0^\omega, 1^\omega\}$}{6mm} &
\tabularblocksingleton{\includegraphics[viewport=50 0 235 90,height=10.8mm]{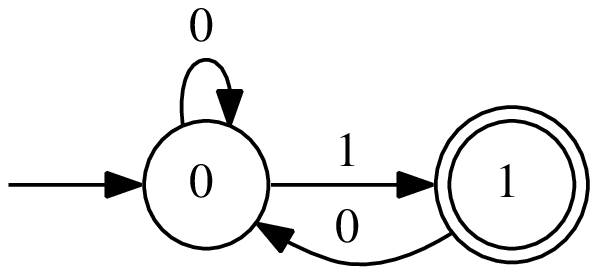}}\\
\end{tabular}\\
\caption{Example calculations of candidate automata from sets of words
\label{exampleautomatafromwords}
}
\end{table}
Table~\ref{exampleautomatafromwords} shows some calculations of
candidate automata from sets $G$ and $B$ obtained in this way.
We remark that even though in the right column the sets $G$ and $B$ are
swapped, the resulting automata are not complementary as for example
neither automaton accepts $01^\omega$.

%We remark that the needed count of wordexamples for forcing a given
%automaton size vary.

%\newpage

\subsection{Minimization algorithm}
\label{algorithm}
This part describes the minimization algorithm; for a given automaton
$A$ find an automaton $A_\tmin$ such that $A_\tmin$ is equivalent
to $A$ and no automaton with fewer states is equivalent to $A$.

\begin{itemize}
\item
Step 1: Choose sets of ultimately periodic words $G$ and $B$.
One may use empty sets; any sets of words such
that $G\subseteq L(A) \subseteq \overline B$ are adequate.

%As required by the candidate finder the algorithm starts choosing
%sets of ultimately periodic words $G$ and $B$. One may use empty sets;
%more efficient is the usage of some small (respective their
%representation) word examples classified by the automaton.
%What words
%are useful are subject of benchmarks; at time
%of writing this text $a^\omega$, $ab^\omega$ and $(ab)^\omega$ for
%all letters $a\not=b$ have turned out to be reasonable.
\item
Step 2: Use the candidate finder to gain some automaton $A'$ out of
$G$ and $B$ with minimal number of states such that
$G\subseteq L(A') \subseteq \overline B$.
\item
Step 3: If $L(A)=L(A')$ then $A'$ is returned as minimal automaton;
in the opposing case choose some counterexample $uv^\omega$
and expand $G$ or $B$ with it. Now resume at step 2
with the bigger sets.
\end{itemize}

This algorithm terminates as the sets $G$ and $B$ hinder any automaton
occured once to occur again. Furthermore there are only finitely many
automata smaller than $A$ so after finitely many steps $A$ would be
returned if no smaller equivalent automaton could be found.

Furthermore the automaton returned has to be equivalent to $A$ as this
is checked before returning the automaton. It is furthermore minimal
as no smaller automaton can separate $G$ and $B$ but every automaton
equivalent to $A$ does so.

\subsection{Implementation}
We have implemented the algorithm in Ocaml, Minisat2 \cite{Minisatseite}
is used as SAT solver\footnote{A download of the program
is available under \tt http://www2.tcs.ifi.lmu.de/\textasciitilde barths/nbamin.html}.

\begin{figure}[h]
\includegraphics[viewport=40 45 670 570,height=80mm]{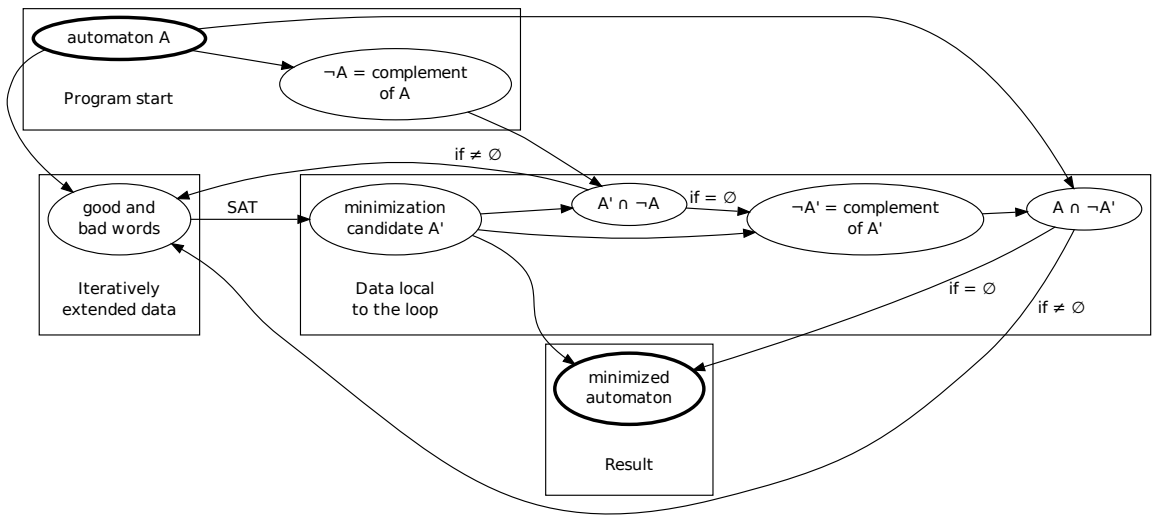}
\caption{Main data flow for minimization\label{datenfluss}}
\end{figure}

Figure~\ref{datenfluss} displays the main data flow while 
Figure~\ref{pseudocode} summarises the complete algorithm in
pseudocode.

\begin{figure}[h]
$ $\\
\tt
00:A = automaton to be minimized;\\
01:negA = complement A;\\
02:G = B = \{\}; (* Sets of good and bad words. *)\\
03:n = 1;\\
04:loop beginning\\
05:\hspace{2.435ex}try A' = NBA-from-solution (SAT-solver (SAT-expression G B n))\\
06:\hspace{4.87ex}failure -> (* No automaton with n states could be found. *)\\
07:\hspace{7.305ex}n := n + 1;\\
08:\hspace{7.305ex}back to loop beginning;\\
09:\hspace{4.87ex}success -> (* A' is candidate for minimized automaton. *)\\
10:\hspace{7.305ex}xB = intersect A' negA;\\
11:\hspace{7.305ex}if xB nonempty\\
12:\hspace{9.74ex}B := B $\cup$ \{onewordfrom xB\}; (* new bad word. *)\\
13:\hspace{9.74ex}back to loop beginning;\\
14:\hspace{7.305ex}else (* L(A') $\subseteq$ L(A) *)\\
15:\hspace{9.74ex}negA' = complement A';\\
16:\hspace{9.74ex}xG = intersect A negA';\\
17:\hspace{9.74ex}if xG nonempty\\
18:\hspace{12.175ex}G := G $\cup$ \{onewordfrom xG\}; (* new good word. *)\\
19:\hspace{12.175ex}back to loop beginning;\\
20:\hspace{9.74ex}else (* L(A) $\subseteq$ L(A') *)\\
21:\hspace{12.175ex}return A'.\\
\caption{Pseudo code for the complete algorithm\label{pseudocode}}
\end{figure}

\paragraph{Calculating example words.}
Counterexamples to $L(A)\stackrel{?}{=}L(A')$ are obtained as words in
the language of NBA $B$ or $C$ which are constructed from $A$ and $A'$ such
that $L(B) = L(A)\backslash L(A')$ and $L(C) = L(A')\backslash L(A)$, thus
finding a word in $L(B)$ or $L(C)$ results in a word
in $L(B) \cup L(C) = L(A) \mathrel{\triangle} L(A')$ where $\triangle$ denotes symmetric difference.
We now describe how to decide
whether for arbitrary NBA $D$ we have $L(D)\neq\emptyset$ and in the
affirmative case how to construct an ultimately periodic
word $uv^\omega\in L(D)$.

We begin by calculating the strongly connected components of $D$
by some linear algorithm, in
our case Kosaraju's
algorithm \cite{kosarajusalgorithm}. Subsequently, we choose
a final state $i$
in a strongly connected component of size at least two or that has a transition
to itself and can be reached from the starting state with some finite
word $u$. There is a path from $i$ to $i$ with some nonempty word $v$ as
$i$ has a transition to itself or lies in a strongly connected component
of size at least two.

From this construction we then know that $uv^\omega \in L(D)$. We further
try to reduce the lengths of $u$ and $v$ by favoring small strongly connected
components that are close to the starting state and by further reducing
the size of $u$ making use of the identity $xy(ly)^\omega=x(yl)^\omega$
where applicable.

\paragraph{Complementation.}
To test for equivalence we need to repeatedly complement
the candidate automata $A'$ as well as the input automaton $A$ itself.
Thus, complementation forms an important component of our algorithm
and the choice of the right algorithm as well as its implementation
will be crucial.

As suggested by \cite{stateofbuechicomp} complementation of NBA by
transforming them into deterministic parity automata (DPA), complementing
them and transform them back to NBA is preferable. Thus this
procedure is used here and leads indeed to small runtimes for that
part of the algorithm. For transformation of NBA to DPA the algorithm
of Safra enhanced by Piterman \cite{nbatodpa} is used.

\subsection{Optimizations}
\label{opti}

We used different optimization to improve the runtime of the algorithm.

\paragraph{Complement storage.}
As the complement of the base automaton is used often we calculate it
at the beginning and store it.

\paragraph{First search for bad words.}
As this does not include complementation of an automaton
it is more efficient to search for words in $\overline A\cap A'$
and skip complementation of candidate automaton $A'$ if a
bad word could be found.

\paragraph{Size reduction of NBA.}
We implemented a series of size reducing algorithms for NBA that require
only linear runtime; they are applied on all intermediate automata and
give a notable optimization of runtime. The used algorithm include
\begin{itemize}
\item Drop unreachable states
\item Drop states where the automaton gets stuck
\item Use a heuristic to detect some states from where all words are accepted. Merge them to one universal state and drop all outgoing transitions
\item Drop transitions that could otherwise have been used to reach that universal state
\end{itemize}

\paragraph{Stop if no smaller automaton found.}

If no smaller automaton was found we have proven minimality and can
return the base automaton.

\paragraph{Choose start words.}

For the needed sets of good and bad words some short (respective
their representation) words are chosen. This does not only reduce
the number of needed calls of the automaton finder but also reduces
the runtime for the single calls of the SAT solver at least if
there are not too many short words in it. How many example words
are useful changes with the introduction of other optimizations and
is adapted by benchmarks from time to time. Currently
the words $a^\omega$, $ab^\omega$, $(ab)^\omega$, $a(ab)^\omega$ for
all different letters $a$ and $b$ as well as $w^\omega$ (where $w$
contains every letter exactly once) are used.

\paragraph{Extra knowledge for the SAT expression.}

We can gain some knowledge out of the automaton to minimize and
include it into the SAT expression. For example if no word starts
with the letter $a$ we know that there is no transition from
the starting state to any states with label $a$.

Giving an order to the states does also gain some speed. This
technique is known as symmetry breaking and is also used.

\subsection{Asymptotic runtime}
Let $\text{SAT}(n)$ be the runtime of a SAT solver on an input of at most $n$
variables and clauses. Let $C(n)$ be the time required to complement
an NBA with at most $n$ states and $c(n)$ the size of this complement
automaton.

The runtime of our minimiser on an automaton of size $N$ with minimal
automaton of size $n$ whose complement has already been computed can
then be summarised by:
\[
O(I \cdot (c(n) \cdot N + n \cdot c(N) +C(n)+SAT(O(I \cdot n^3))))
\]
where  $I$ is the number
of iterations of our algorithm. Obviously, $I=2^{O(n)}$, but in
practice, $I$ is much smaller than this bound.

The factor $I$ in the $SAT$ expression comes from the linear
dependency of the SAT formulas on the number of example words.

Additionally, if the complement of the input automaton is
already known the runtime depends only linearly on the size
of the automaton for different automata describing the
same language.

\section{Experimental results}
\label{experimental}
As said in the previous chapter
%The first experimental result is that
the runtime for minimization
depends much more on the size of the found minimal automaton than on
the size of the original automaton.

Most given runtimes were measured on the same machine;
2300 MHz, Quad-Core AMD Opteron(tm) Processor 8356;
for each calculation one core was used. Vague given
runtimes were calculated on slower machines.

If the minimal automaton is small enough and the complementation of
the automaton is fast enough even large automata can be minimized;
for example we could find some randomly generated automata
with 40 to 100 states whose minimal
equivalent automata of size up to 5 could be found in some minutes; to ensure
that this is the merit of the minimizer we ensured that the heuristic
pre-minimizer could not reduce the size of the original automaton.

\begin{figure}[h]
\begin{tabular}{ll}
\begin{tabular}{l}
$(G(q \lor F G p) \land G ( r \lor F G \neg p)) \lor G q \lor G p
$\\
\includegraphics[height=40mm]{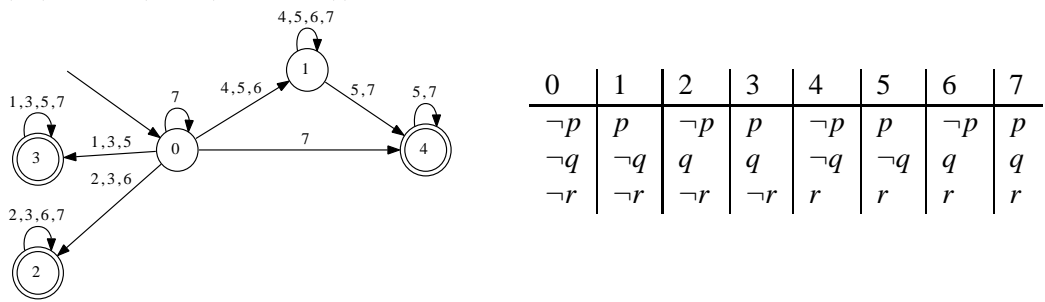}
\end{tabular} &
\begin{tabular}{l|l|l|l|l|l|l|l}
0 & 1 & 2 & 3 & 4 & 5 & 6 & 7\\
\hline
$\neg p$ & $p$ & $\neg p$ & $p$ & $\neg p$ & $p$ & $\neg p$ & $p$\\
$\neg q$ & $\neg q$ & $q$ & $q$ & $\neg q$ & $\neg q$ & $q$ & $q$\\
$\neg r$ & $\neg r$ & $\neg r$ & $\neg r$ & $r$ & $r$ & $r$ & $r$
\end{tabular}
\end{tabular}

\caption{\label{ltlminimize}LTL formula together with its minimal automaton and the boolean value to alphabet translation table}
\end{figure}

\begin{table}[b]
\centering
\begin{tabular}{l|l|l|l|l|l}
Resulting size & count & average time & 10\%-decile & median & 90\%-decile time\\
\hline
\hline
1 & 245 & $9.71\cdot 10^{-2}\, \text s$ & $5.66\cdot 10^{-3}\, \text s$ & $7.65\cdot 10^{-3}\, \text s$ & $2.85\cdot 10^{-1}\, \text s$\\
\hline
2 & 179 & $2.98\cdot 10^{-1}\, \text s$ & $4.08\cdot 10^{-2}\, \text s$ & $3.00\cdot 10^{-1}\, \text s$ & $5.45\cdot 10^{-1}\, \text s$\\
\hline
3 & 76 & $2.04\, \text s$ & $1.50\cdot 10^{-1}\, \text s$ & $1.96\, \text s$ & $4.02\, \text s$\\
\hline
4 & 80 & $9.82\, \text s$ & $3.33\, \text s$ & $9.13\, \text s$ & $2.08\cdot 10^{1}\, \text s$\\
\hline
5 & 66 & $4.30\cdot 10^{1}\, \text s$ & $1.46\cdot 10^{1}\, \text s$ & $3.78\cdot 10^{1}\, \text s$ & $8.47\cdot 10^{1}\, \text s$\\
\hline
6 & 53 & $7.77\cdot 10^{2}\, \text s$ & $1.04\cdot 10^{2}\, \text s$ & $2.96\cdot 10^{2}\, \text s$ & $1.50\cdot 10^{3}\, \text s$\\
\hline
7 & 38 & $7.61\cdot 10^{3}\, \text s$ & $4.16\cdot 10^{2}\, \text s$ & $3.14\cdot 10^{3}\, \text s$ & $2.24\cdot 10^{4}\, \text s$\\
\hline
8 & 2 & $2.01\cdot 10^{4}\, \text s$ & $7.78\cdot 10^{3}\, \text s$ & --- & $3.24\cdot 10^{4}\, \text s$\\
\hline
Abnormal termination & 199
\end{tabular}
\caption{Measured minimization times for automata of starting size 10.\label{minitime}}
\end{table}

Furthermore we used a simple LTL to NBA translator that intentionaly
does not optimize very well, just
%further than
using our heuristic minimizer for the intermediate
steps of the construction.
%This leads to by far not minimal NBA.
Nevertheless we can minimize them if the minimal size is not too
big. A formula (taken from \cite{spin2010}; details of the experimental
evaluation, formula 1.22) that lead to an automaton
of size 157 and could be minimized in half an hour is shown in
Figure~\ref{ltlminimize} together with its minimal automaton.
Remark that \cite{spin2010} used a partial minimizer on an 8 state
version of this automaton and only could find a 6 state automaton
representing this formula; we could find a 5 state automaton
without using a well pre-minimized NBA.

Speed measurement is given in Table~\ref{minitime}. Random
NBA with $10$ states and alphabet size two were generated. States
are final with
probability $0.5$; for every two states $i,j$ and letter $a$ there
is a transition from $i$ to $j$ with probability $0.15$. If there
are unreachable states or states from where no word can be accepted
the automaton is skipped.
Abnormal termination means out of memory or timeout (12h). When
starting with 7-state automata the table looks similar but has
no abnormal termination; it does not give additional information
about the runtime and is hence skipped.

Table~\ref{nbaut} shows the minimization results for complement
automata of all automata with small size; as complementation can
result in exponential blowup this needed minimizations of automata
of bigger sizes.

\begin{table}[h]
\centering
\begin{tabular}{l|l|l}
%number of states & 2 & 2\\% & 3\\
%$|\Sigma|$ & 2 & 3\\% & 2\\
%Number of different automata & 768 & 12288\\% & 786432\\
$|\text{states}|$ / $|\Sigma|$ / \#different automata & 2/2/768 & 2/3/12288\\
\hline
\hline
\#reducing to size 1 & 478 & 4404\\% & ...\\
\#reducing to size 2 & 290 & 7884\\% & ...\\
%\#reducing to size 3 & --- & --- & ...\\
\hline
\#minimal complement size 1 & 372 & 2850\\% & ---\\
\#minimal complement size 2 & 206 & 2754\\% & ---\\
\#minimal complement size 3 & 134 & 3024\\% & ---\\
\#minimal complement size 4 & 40 & 2429\\% & ---\\
\#minimal complement size 5 & 16 & 1039\\% & ---\\
\#minimal complement size 6 & --- & 180\\% & ---\\
\#minimal complement size 7 & --- & 12\\% & ---\\
%\hline
%maximal runtime ($t_{\text{max}}$) & 0.3 sec & 1.1 sec\\% & ...\\
%average runtime ($t_{\text{avg}}$) & 0.03 sec & 0.1 sec\\% & ...\\
%standard deviation of runtime ($t_{\text{stdev}}$) & 0.04 sec & 0.1 sec\\% & ...\\
%\hline
%$t_{\text{max}}$ for complement minimization & 53 min & ---\\% & ---\\
%$t_{\text{avg}}$ for complement minimization & 9 sec & ---\\% & ---\\
%$t_{\text{stdev}}$ for complement minimization & 2 min & ---\\% & ---\\ % 135 sec
\end{tabular}
\caption{\label{nbaut}Complete sampling of automata with small sizes}
\end{table}

\begin{figure}[h]
\begin{tabular}{ll}
\includegraphics[height=20mm]{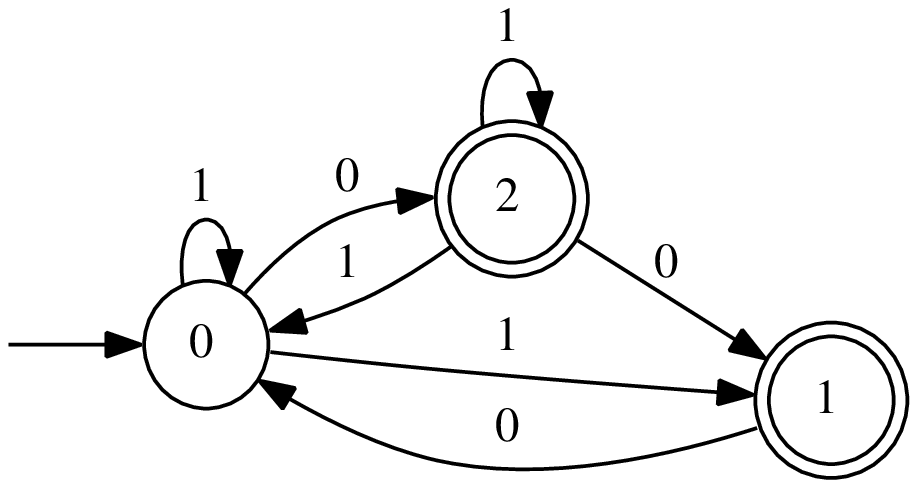} &
\includegraphics[height=30mm]{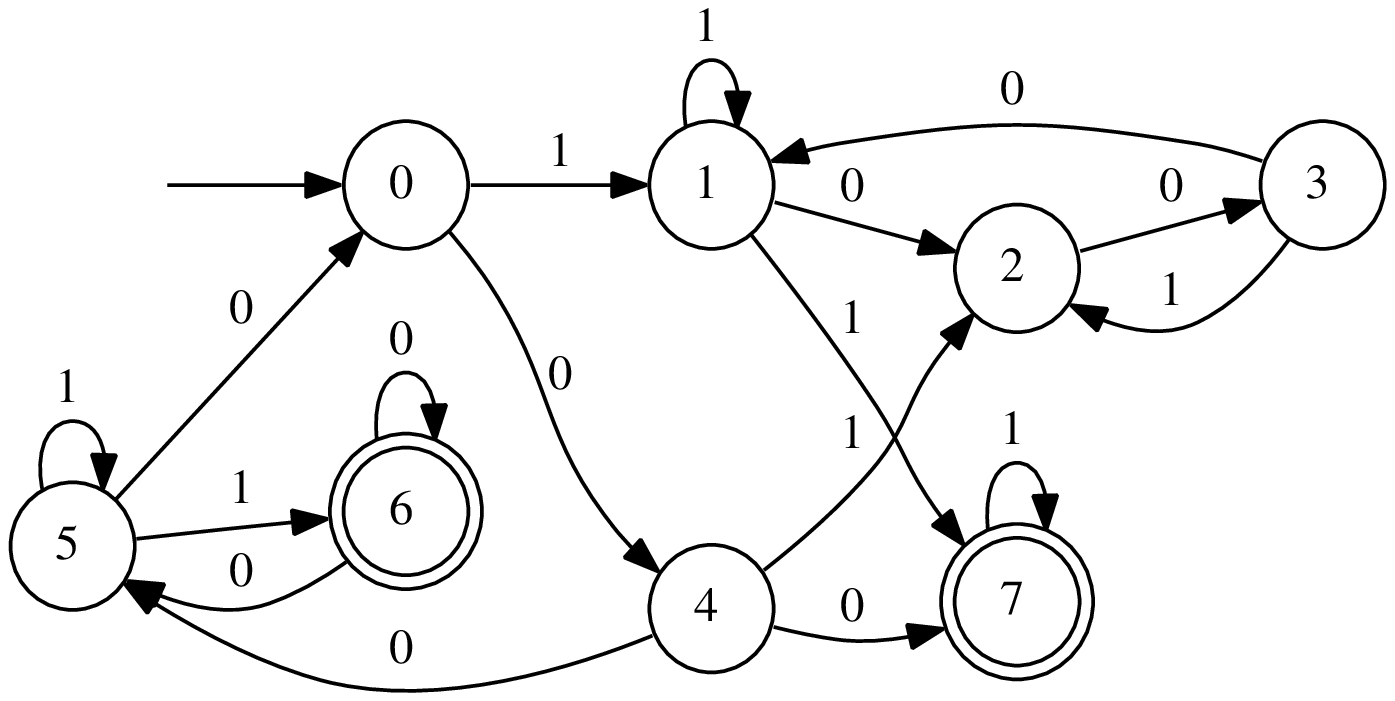}\\
(a) & (c)
\end{tabular}\\
\begin{tabular}{l}
\includegraphics[viewport=45 45 980 455,height=51mm]{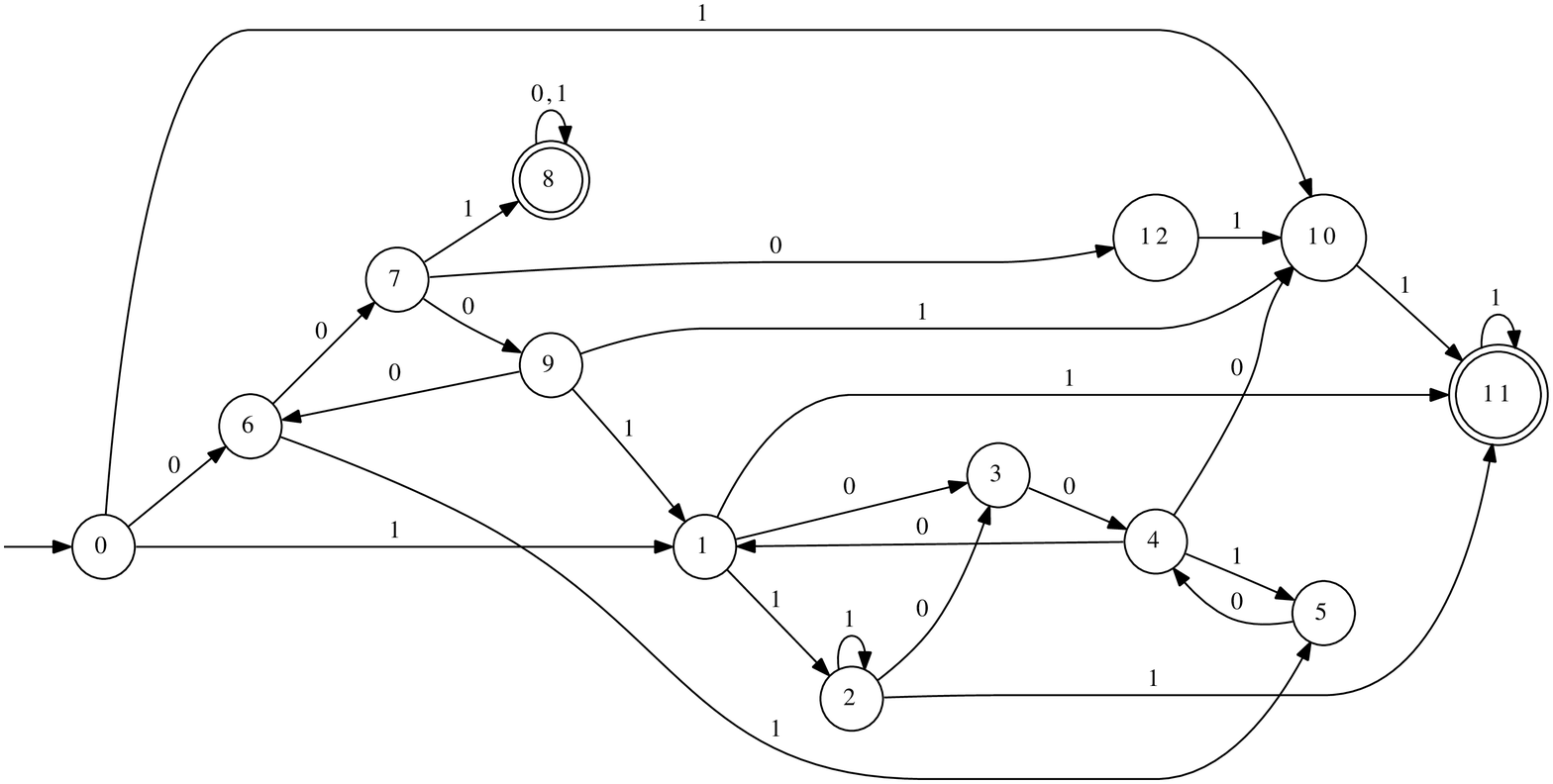}\\
(b)
\end{tabular}
\caption{\label{big32comp}Automaton with 3 states and 2-letter-alphabet (a) with minimal complement size 8, its complement from the complementation (b) and its minimal complement automaton (c)}
\end{figure}

%\newpage

Having all these automata minimized one can now be sure that no automaton
with two states and two letter alphabet needs more than 5 states for its
complement. For three letter alphabet this limit is increased to 7 states.
Only two (up to alphabet permutation) automata reach this limit.

Work is in progress to minimize all complements of automata with three states
and two letter alphabet; an automaton with minimal complement of size 8 was
found hereby;
% this automaton, its complement out of the complementation
%and its minimal complement automaton are shown
it is presented in Figure~\ref{big32comp}.

We also run our procedure on several instances of Michel's automata $M_n$ over the alphabet
%Beside random examples we ran the minimization and complement
%minimization of Michel's automata $M_n$ over the alphabet
$\Sigma=\{0,\dots,n\}$ and with $n+1$ states \cite{Michel88} which
were introduced to establish an $n!$ lower bound for complementation
of NBA. Indeed, Michel has shown that no NBA with fewer than $n!$
states can recognize the complement of $L(M_n)$.

\begin{figure}[h]
\centering
\begin{tabular}{ll}
\includegraphics[viewport=85 30 265 125,clip,height=13mm]{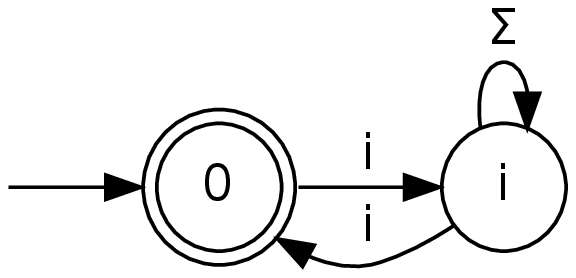} &
\includegraphics[viewport=45 0 405 185,clip,height=25mm]{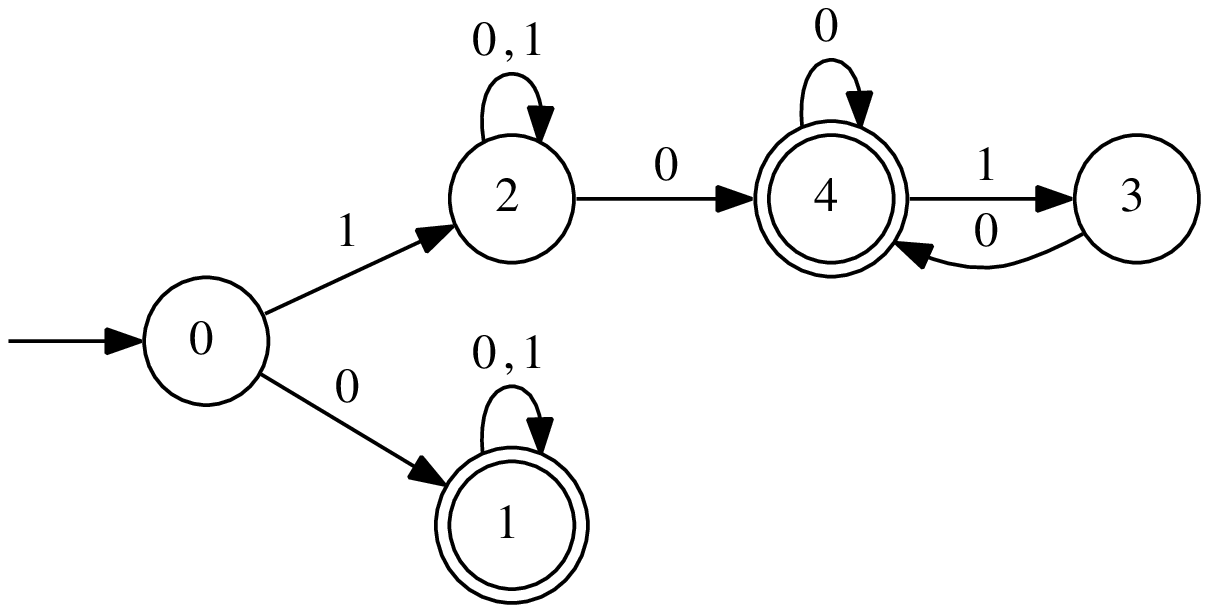}
\end{tabular}
\caption{Michel automaton schematically and minimal complement of $M_1$
\label{michelfigure}}
\end{figure}

The automata $M_n$ are given schematically on the left side in
Figure~\ref{michelfigure} where $i$ represents a number
in $\{1,\dots,n\}$, so $i\neq 0$.

We needed under a minute to compute the minimal complement
of $M_1$; for $M_2$ we could prove that at least $7$ states
are needed to represent it while the full minimization process
timed out.

The minimality of $M_n$ for $1 \leq n \leq 5$ could be proven as well.

\begin{figure}[h]
\centering
\begin{tabular}{lllll}
\includegraphics[viewport=50 0 355 260,height=3.12cm]{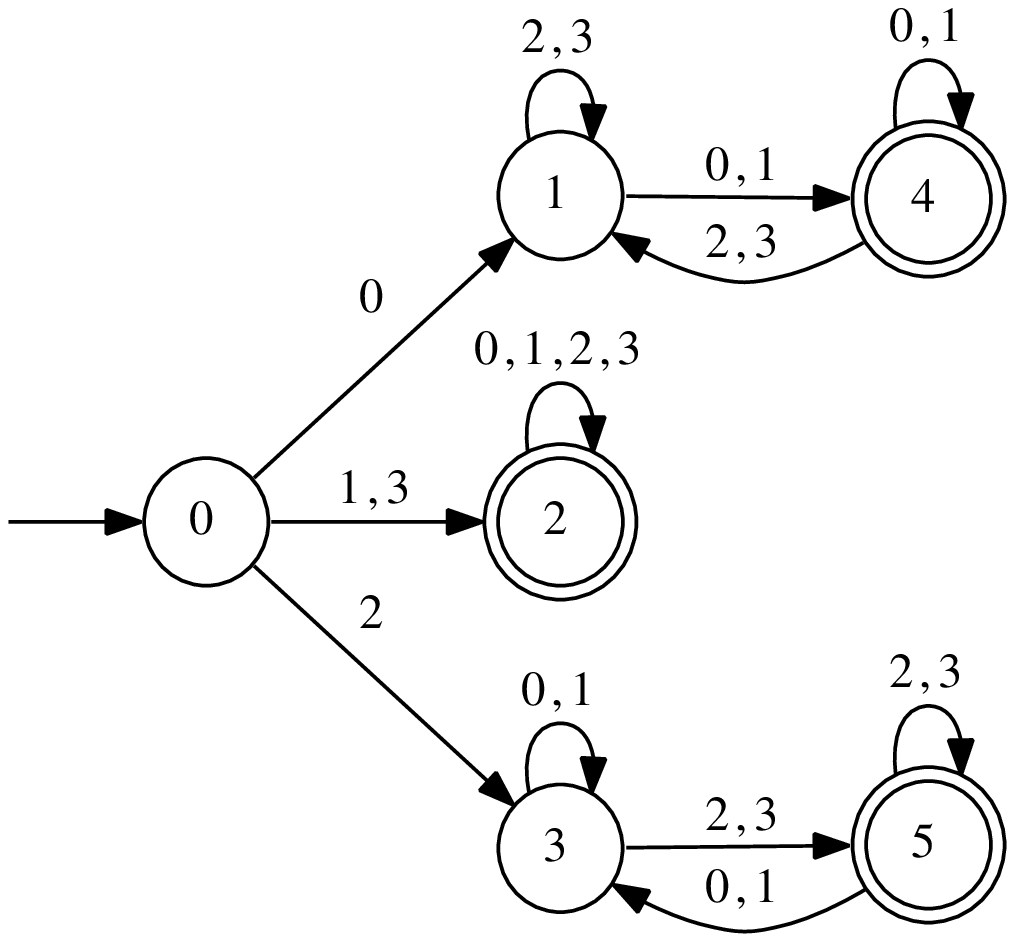} & ~ &
\includegraphics[viewport=50 0 360 195,height=2.34cm]{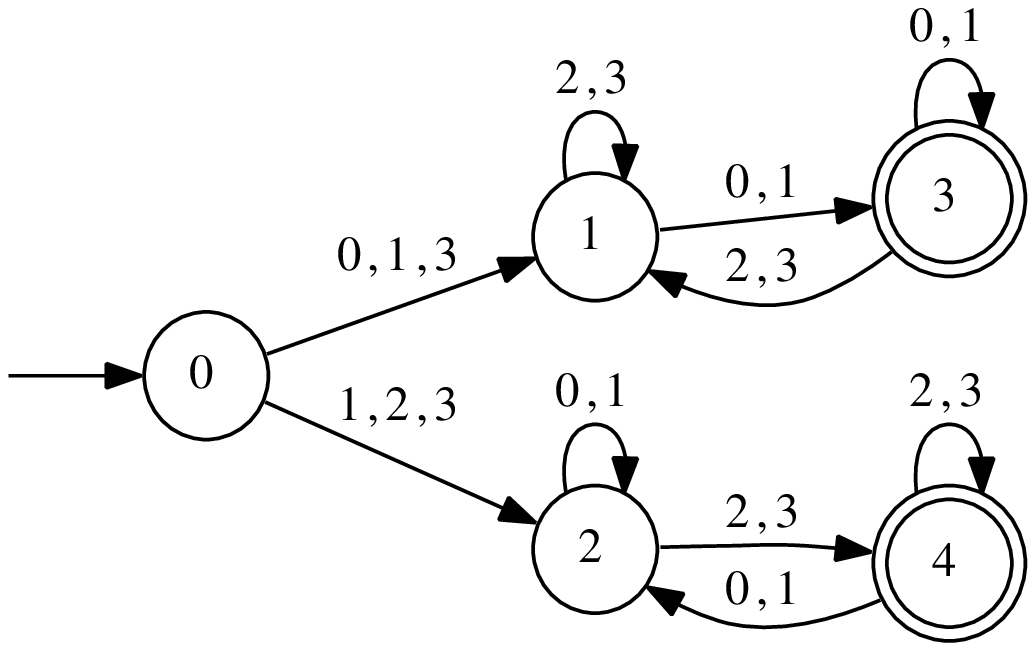} & ~ &
\includegraphics[viewport=55 0 365 260,height=3.12cm]{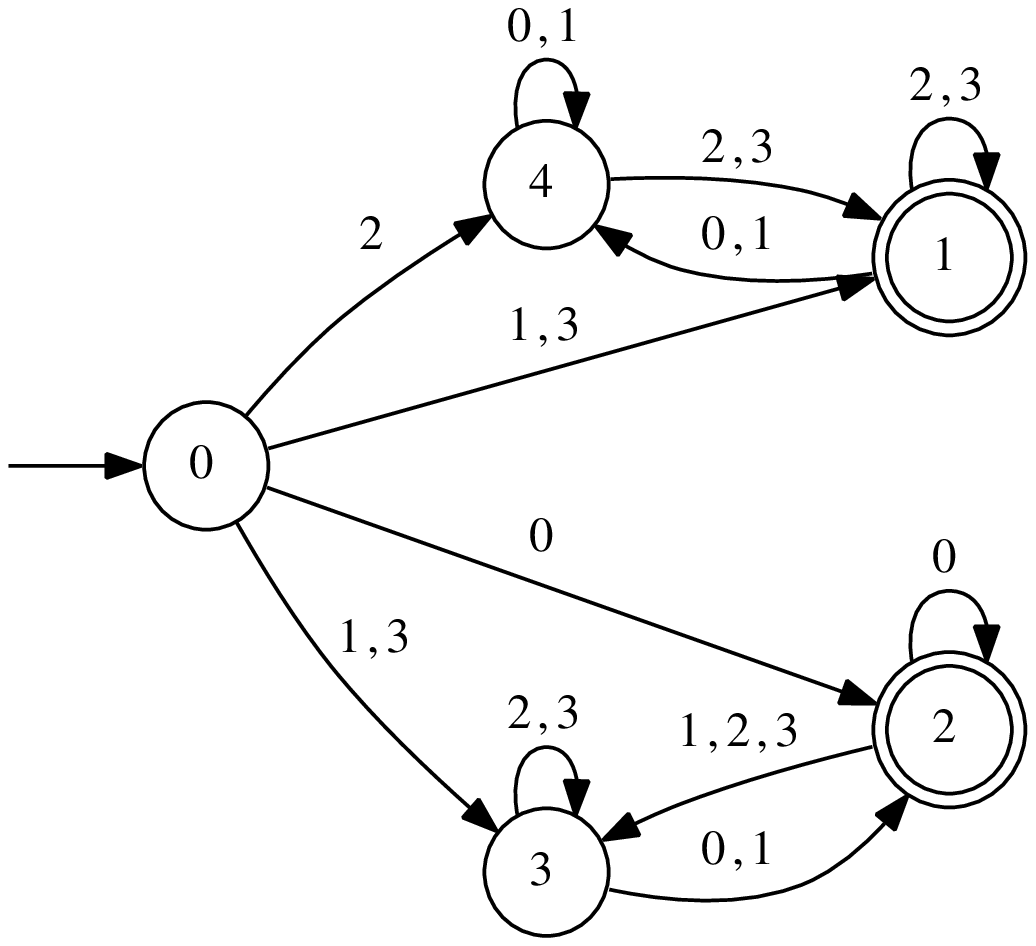} \\ % 295
(a) & ~ & (b) & ~ & (minimized)
\end{tabular}
\caption{(a), (b): Automata shown in \cite{spin2010} page 15 Figure 5;
alphabet was chosen with letter $0$ for $(\neg p, \neg r)$, $1$ for $(\neg p, r)$, $2$ for
$(p, \neg r)$, $3$ for $(p, r)$;
%(a) and (b) are language equivalent but not bounded language equivalent.
(minimized) is the result from our algorithm
\label{spin2010figure5}
}
\end{figure}

Another calculated minimization example was taken from \cite{spin2010},
a paper describing a minimization algorithm of NBA wherein a stronger form of
equivalence, so-called bounded language equivalence, is used.
It is presented in Figure~\ref{spin2010figure5}.
The automata shown are language equivalent, but not bounded
language equivalent. As result a minimizer based on bounded language equivalence
could not find (b) as minimal automaton for (a).

Our complete minimizer could minimize the 6 state, 4 letter automaton
(a) under a minute, leading to the
%and complement minimize it in 38 seconds
result shown in Figure~\ref{spin2010figure5} (minimized).
It did not find the automaton (b) from
Figure~\ref{spin2010figure5}, but instead another language equivalent
but not bounded language equivalent automaton of the same size 5.

%\newpage

\section{Conclusion}
We have established the first global minimization algorithm
for arbitrary nondeterministic B\"uchi automata.
Previous algorithms were either restricted to special
classes of B\"uchi automata or computed the automaton with the least
number of states among those reachable from a given one by several
optimization steps.

%Compared with a brute force
%algorithm that enumerates B\"uchi automata and checks each one for
%equivalence with the given one. We remark that the number of B\"uchi
%automata with 5 states and alphabet size 2 exceeds $10^{16}$ and for
%any one of these a costly equivalence test would have to be
%performed. Even assuming that any one of those could be performed in
%1ms we would still need more than 100000 years to explore them all so as to
%ascertain that a given automaton with 6 states is indeed minimal.

% The algorithm is fairly slow and not least due to the inherent
% complexity of the problem does not scale to large examples.

Despite the exponential worst-case running time of our algorithm we
succeeded in applying it to several nontrivial automata with an
acceptable runtime and in this way established previously unknown
facts. Several people asked for a comparison with a naive brute
force enumeration of all B\"uchi automata. We note here that already
the number of automata with 5 states and alphabet size 2 exceeds
$10^{16}$ and for every one of these a costly equivalence test would
have to be performed which means that this procedure is infeasible for
input automata with six or more states.

Of course, we did not establish a new upper bound of complexity with
our algorithm but this was not to be expected as minimization of
B\"uchi automata is PSPACE-complete. We also note that it has become
common practice with good practical results to develop and use
algorithms with exponential worst case runtime, e.g.\ SAT-solvers, or
model checkers for LTL.

In particular we were able to assert that no B\"uchi automaton with
two states and alphabet size two has a minimal complement automaton
with more than five states and that the minimal complement automaton
of Michel~\cite{Michel88} for alphabet size two achieves this
bound. With the brute force enumeration such result would have been
impossible to obtain even assuming some heuristic strategies to rule
out candidates.

The implementation of the relatively straightforward optimizations
described in Section~\ref{opti} each produced considerable speedups;
we thus hope that further relatively easy optimizations would allow us
to push the limit of feasibility further out and make more
applications accessible to our method.
For all tested automata over a size of 4 for the minimal automaton
over 50\% of computational time went into the SAT-solver, most
times over 99\% of time is used here so further optimization focuses here.

We were asked to what extent our algorithm is able to produce
certificates of the asserted minimality of its output. Since
minimization is PSPACE complete we cannot in general expect
polynomially sized certificates unless NP=PSPACE. However, we can
remark here that the final sets of good ($G$) and bad words ($B$)
together with the purported size $n$ of the minimal automaton can
serve as a ceritificate of sorts in the following way. An opponent who
is not convinced of the asserted result can first check that
$G\subseteq L$ and $B\subseteq\overline{L}$ where $L$ is the language
of the original automaton to be minimized. Thereafter, they could
construct the SAT formula searching for an automaton of size $n-1$
whose language $L'$ satisfies $G\subseteq L'\subseteq
\overline{B}$. Alternatively, we could provide a corresponding
resolution proof. While potentially large and difficult to check, these
certificates are considerably more concise and intuitively valid
than the always-open
fallback option of a complete trace of a run of the algorithm.

In particular, in
automata-based software model checking
\cite{Holzmann:2003:SMC:1405716} one must check that all runs of a
program are accepted by an often small policy automaton. Minimizing
the latter might result in considerable gains if it is used repeatedly
on many different programs. Consider e.g.\ that the automaton
represents some publically advertised security policy or even a
standard.

We also anticipate possible usages of our algorithm as a tool for
research into B\"uchi automata and teaching thereof. It could for
example be used to early refute hypotheses about the strength of
minimization heuristics yet to be invented. Notice here that our
algorithm was able to further minimize the automaton from
\cite{spin2010}.

Finally, it will also be interesting to apply our SAT-based search to
other instances of ``minimally adequate teachers'' for
$\omega$-regular languages, in particular the ones arising from
compositional verification \cite{NFM-2010}.

%\newpage
\bibliographystyle{eptcsalpha}
%\pdfbookmark{\refname}{\refname}
%{\small
\bibliography{minibuechi}
%}

\end{document}